\documentclass[aps,pra,twocolumn,floatfix]{revtex4-1}
\usepackage[dvips]{graphicx}
\usepackage{ulem}
\usepackage{cancel}

\usepackage[dvipdfm,pdfstartview=FitH]{hyperref}
\usepackage{mathrsfs}
\usepackage{float}
\usepackage{graphicx}
\usepackage[english]{babel}
\usepackage{amsmath}
\usepackage{amssymb}
\usepackage{CJK}
\usepackage{slashed}
\usepackage[dvipdfm,pdfstartview=FitH]{hyperref}
\usepackage{bm}% bold math

\newcommand{\be}{\begin{eqnarray}}
\newcommand{\ee}{\end{eqnarray}}
\newcommand{\ba}{\begin{eqnarray}}
\newcommand{\ea}{\end{eqnarray}}

\newcommand{\Ek}{E_{\mathbf{k}}}

\newcommand{\vq}{\mathbf{q}}
\newcommand{\sumk}{\sum_{\mathbf{k}}}

\newcommand{\xik}{\xi_{\mathbf{k}}}

\newcommand{\nn}{\nonumber}

\allowdisplaybreaks
\begin{document}

\title{Dynamic density structure factor of a unitary Fermi gas at finite temperature}

\author{Hao Guo$^1$, Yan He$^2$, and Lianyi He$^3$}
\affiliation{$^1$Department of Physics, Southeast University, Nanjing, Jiangsu 211189, China}

\affiliation{$^2$College of Physical Science and Technology,
Sichuan University, Chengdu, Sichuan 610064, China}\email{heyan_ctp@scu.edu.cn}

\affiliation{$^3$Department of Physics and State Key Laboratory of Low-Dimensional Quantum Physics, Tsinghua University, Beijing 100084, China}

%\date{\today}

\begin{abstract}
We present a theoretical investigation of the dynamic density structure factor of a strongly interacting Fermi gas near a Feshbach resonance at finite temperature. The study is based on a gauge invariant linear response theory.  The theory is consistent with a diagrammatic approach for the equilibrium state taking into account the pair fluctuation effects and respects some important restrictions like the $f$-sum rule. Our numerical results
show that the dynamic density structure factor at large incoming momentum and at half recoil frequency has a qualitatively similar behavior as the order parameter, which can signify the appearance of the condensate.
This qualitatively agrees with the recent Bragg spectroscopy experiment results. We also present the results at small incoming momentum.

\end{abstract}

\maketitle

\section{Introduction}

Ultra-cold Fermi gas has been the focus of a lot of research investigations due to its highly controllable attractive interaction\cite{Bloch-RMP,NSR,GriffinPRL02,Ourreview,HuHuiPRA05,KinastPRL05}. Comparing to other condensed matter systems, it is very difficult to establish the appearance of condensate or phase coherence in cold fermi gases, because of the lack of transport measurements. A decade ago, the superfluidity of Fermi gases has been proved experimentally by the observation of the vortex lattices\cite{Zwierlein-vortex}. But this method still requires a fast sweep of the attractive interaction between the fermions to the deep Bose-Einstein condensation (BEC) limit, in order to see the density depletion in the vortex core. Recently, an alternative approach based on measuring the collective (Nambu-Goldstone) mode though the Bragg spectroscopy has been carried out experimentally to establish the phase coherence of unitary Fermi gases \cite{Vale14}. Due to the inhomogeneity of the clouds of Fermi gases, the integrated Bragg spectroscopy is difficult to extract useful information to indicate the existence of the condensate. This issue has been overcome by a method invented in Ref.\cite{Vale14}, which makes the local observation of Bragg spectroscopy possible.

A lot of information of the many-body stsytem can be inferred from the dynamic structure factor, such as the excitations related to the pair breaking and quasiparticle scattering. Importantly, the Nambu-Goldstone mode due to the breaking of U(1) symmetry in the superfluid/superconducting phase can also be deduced when the momentum transfer is small, which can help to judge the onset of the ordered phase.
A theoretical account of Bragg spectroscopy requires a consistent calculation of the density-density response function or the dynamical density structure factor (susceptibility)\cite{Stringari06,HaoPRL10}. The word ``consistent'' here means that one must take into account of the contributions from condensate, pair fluctuation and collective modes in a gauge invariant fashion such that certain conservation laws must be respected. A naive tree-level calculation of the density response function cannot even produce the correct locations of poles associated with the collective modes. Due to the strong correlations occurred in the unitary limit, the BCS-BEC crossover theory of unitary Fermi gases has to employ certain approximations to compute various thermodynamic properties\cite{Randeria97,Haussmann}. It is a challenging problem to maintain the gauge invariance of the linear response theory when extensions of these approximations are included. The ``consistency'' of the theory can be thought of as an important constraint since no exact solution is known for strongly correlated Fermi gas.

Important progresses have been made in the theoretical framework of the structure factor. Early development of quantum Monte Carlo (QMC) approach was focused on low temperature simulation\cite{CombescotEPL06}, and later applied to inhomogeneous unitary Fermi gas at finite temperature \cite{GandolfiPRA14}. Diagrammatic technique was applied in Refs.\cite{Combescot06,HaoPRL10,StrinatiPRL12}, and the random-phase approach (RPA) was developed in Ref.\cite{HuHuiPRA10}. The latter was later generalized to the RPA on top of the superfluid local density approximation (SLDA-RPA) approach.

In this paper, we adopt an $G_0G$ pair fluctuation theory to compute the density response function of unitary Fermi gases. We show that a gauge invariant linear response theory of which the approximation exactly matches the approximation used in the thermodynamic calculation can be constructed. In this way, one can show that the resulting response function will satisfy the current conservation and various sum rules exactly. Our theory also has the advantages of easy implementation in the numerical calculations.

This paper is organized as follows.
We first brief introduce $G_0G$ pair fluctuation theory both below and above $T_c$ in section \ref{sec-G0G}. Then we show how to construct a gauge invariant linear response theory in section \ref{sec-th}. In section \ref{sec-num}, we present the numerical results of our theory and compare with experiments.

\section{A brief introduction to the Pair Fluctuation theory}
\label{sec-G0G}
In this paper, our evaluation of dynamical density susceptibility is based on the $t$-matrix theory with ladder diagrams made by one bare and one fully dressed Green's functions, which is known as the ``$G_0G$" theory. This approach is inspired by the early work of Kadanoff and Martin \cite{KM}. A detailed review of this theory can be found in Ref.\cite{Ourreview}. This asymmetric choice of the ladder series is more compatible with the BCS-leggett ground state\cite{KM,Maly1}. Before explaining the main idea of the theory, we first fix some notations as follows.

Assuming that $m$ is the particle mass and $\mu$ is the chemical potential, the general one-particle Green's function is
\begin{equation}
G^{-1}(K)=G^{-1}_{0}(K)-\Sigma(K).
\end{equation}
Here $G_{0}(K)=1/(i\omega_{n} -\xik)$ is the non-interacting fermionic Green's function, $\Sigma(K)$ is the self-energy and $\xik=\varepsilon_{\bf k}-\mu$ is the free fermion dispersion with $\varepsilon_{\bf k}={\bf k}^2/(2m)$. We adopt the following convention: $k_{B}=1$, $\hbar=1$, four-vector $K=(\omega_n, \mathbf{k})$ with $\omega_{n}=(2n+1)\pi\beta$, $n$ is an integer, $\beta=1/T$ and $\sum_{K}=T\sum_{n}\sum_{\mathbf{k}}$.

For the strongly attractive Fermi gas, the pair fluctuations can be treated as virtual non-condensed pairs in equilibrium with the condensate of Cooper pairs. To describe the effects of non-condensed pairs, we introduce the $t$-matrix $t_{\textrm{pg}}(Q)$ which can be thought to be an amputated propagator for non-condensed pairs. Below $T_c$ in the superfluid phase, the self-energy of the $G_0G$ $t$-matrix theory can be decomposed into two parts. Aside from the usual BCS self-energy $\Sigma_{\textrm{sc}}(K)=\Delta^2_{\textrm{sc}} G_0(-K)$ with $\Delta_{\textrm{sc}}$ being the order parameter, there is a pseudogap self-energy which is dressed by the pair propagator or $t$-matrix as
$
\Sigma_{\textrm{pg}}(K)=\sum_Q t_{\textrm{pg}}(Q)G_0(Q-K)\label{sig}$.
The pair propagator is given by the summation of infinite ladders made of bare and full Green's functions, and its expression is given by
\be
&&t_{\textrm{pg}}(K)=\frac{g}{1+g\chi(K)},\\
&&\chi(K)=\sum_{Q}G_0(K-Q)G(Q).
\ee
Here $g$ is the contact coupling constant, which is related to the $s$-wave scattering length $a_s$ through
\begin{equation}
\frac{1}{g}=-\frac{m}{4\pi a_s}+\sum_{\bf k}\frac{1}{2\varepsilon_{\bf k}}.
\end{equation}
This renormalization relation cancels precisely the ultraviolet divergence in $\chi(K)$.

The Bose-Einstein condensation condition of the non-condensed pairs in this case can be expressed as the vanishing of the ``pair chemical potential". Then the condensation condition is equivalent to the divergence of of the pair propagator at zero momentum $t^{-1}_{\textrm{pg}}(0)=1+g\sum_KG_0(-K)G_(K)=0$ which is just the BCS gap equation. Therefore, the $G_0G$ theory reduces to BCS mean field theory at $T=0$.

There is an undetermined pseudogap self-energy in the $t$-matrix which determines the pseudogap self-energy itself. Therefore, the full $G_0G$ $t$-matrix theory requires a  self-consistent solution to $\Sigma_{\textrm{pg}}$ from a set of coupled integral equations, which is too complicated in practical calculations. One can employ an approximation to simplify the final result. The pair condensation condition, or the Thouless criterion, $t^{-1}_{\textrm{pg}} (0) = 0$, implies that
the main contribution to $\Sigma_{\textrm{pg}}$ comes from the vicinity of $Q = 0$. Thus, one can simplify the convolution to a multiplication
\be\Sigma_{\textrm{pg}}(K)\approx\Big[\sum_Q t_{\textrm{pg}}(Q)\Big]G_0(-K)\equiv-\Delta_{\textrm{pg}}^2G_0(-K).\ee
In this way, $\Sigma_{\textrm{pg}}$ takes the same form as the BCS self-energy, which greatly simplify the numerics and also provides an explicit expression for the pseudogap $\Delta_{\textrm{pg}}$. Hence the total self-energy is given by $
\Sigma(K)=\Sigma_{\textrm{sc}}(K)+\Sigma_{\textrm{pg}}(K)$.
Then energy gap can be decomposed into the superconducting (sc) and the
pseudogap (pg) parts
%\begin{equation}\label{eq:Delta}
$\Delta^{2}=\Delta_{\textrm{sc}}^{2}+\Delta_{\textrm{pg}}^{2}$.
%\end{equation}
Now $T_c$ is determined by the vanishing of the order parameter $\Delta_{\textrm{sc}}$,
 while $T^*$, the onset temperature of pairing, is determined
by the vanishing of the total gap $\Delta$. Below $T_c$, the gap, number
and pseudogap equations can be summarized as
\ba
\frac{1}{g}&=&\sum_{\mathbf{k}}\frac{1-2f(\Ek)}{2\Ek},\label{eq:gap}\\
n&=&\sumk\Big[1-\frac{\xik}{\Ek}+2f(\Ek)\frac{\xik}{\Ek}\Big],\label{eq:neq}\\
\Delta_\textrm{pg}^2&=&a_0^{-1}\sum_{\vq}b(\Omega_{\vq}).
\label{eq:pgeq}
\ea
Here $f(x)=1/(e^{x/T}+1)$ and $b(x)=1/(e^{x/T}-1)$ are Fermi and Bose distribution functions, and $E_{\bf k}=\sqrt{\xik^2+\Delta^2}$. $a_0$ and $\Omega_{\vq}={\bf q}^2/(2M)$ are the pair propagator residue and pair dispersion with an effective mass $M$ respectively, which can be obtained by
the following expansion
\be
t^{-1}_\textrm{pg}(\omega,\vq)=a_0(\omega-\Omega_{\vq}),\quad T<T_c.
\ee
Above $T_c$, there is no condensate and $\Delta_\textrm{sc}=0$. We expect a non-zero pair chemical potential appearing in the following expansion
\be
t^{-1}_\textrm{pg}(\omega,\vq)=a_0(\omega-\Omega_{\vq}+\mu_p),\quad T>T_c.
\ee
Accordingly the gap and pseudogap equations are replaced by
\ba
\frac{1}{g}&=&\sum_{\mathbf{k}}\frac{1-2f(\Ek)}{2\Ek}-a_0\mu_p,\\
\Delta_\textrm{pg}^2&=&a_0^{-1}\sum_{\vq}b(\Omega_{\vq}-\mu_p).
\ea

So far we have summarized the most important ingredients of $G_0G$ pair fluctuation theory. This theory has been applied to compute various thermodynamical quantities in the BCS-BEC crossover of ultra-cold Fermi gases, and agrees with the experiments well\cite{Yanthermo}. A gauge invariant linear response theory based on it must contain the approximations in the same way as how the pair fluctuation effect is included in the self-energy.

\section{Gauge Invariant Linear Response Theory with Pair Fluctuation Effects}
\label{sec-th}

\begin{figure}[t]
\includegraphics[clip,width=0.45\textwidth]{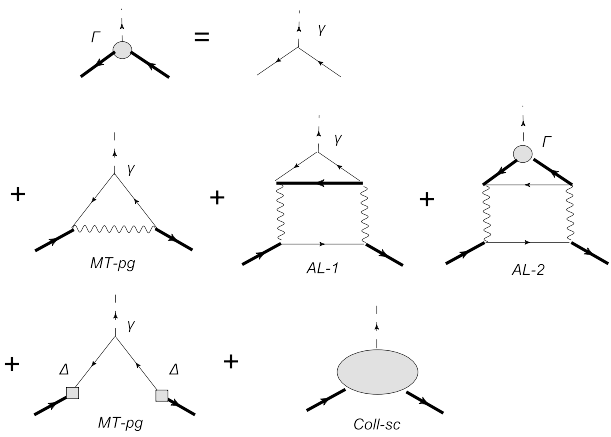}
\caption{A diagrammatic representation of Eq.(\ref{V1}). The thin and thick straight lines correspond to bare and full propagator of a single fermion respectively. The wavy line denotes the pair propagator $t_\textrm{pg}$. The small dot and big circle are the bare and full interaction vertices respectively. The small square is the pairing field. The ellipse is the collective-mode contribution.}\label{MT-AL}
\end{figure}

The Hamiltonian of the ultra-cold Fermi gas has a U(1) symmetry, of which the Noether current $J^\mu$ satisfies the conservation law $\partial_\mu J^\mu=0$. The density-density response function function can be obtained from a linear response theory subjected to an effective external electromagnetic (EM) potential $A_{\mu}$ where $\mu=0,1,2,3$ is vector index of the pseudo Minkowski space with metric $\eta^{\mu\nu}=\textrm{diag}(1,-1,-1,-1)$. The induced EM current is given by the Kubo formalism
$\delta J^{\mu}(Q)=K^{\mu\nu}(Q)A_{\nu}(Q)$, where $Q=(\omega,\vq)$ is the four-momentum of external field.
The EM response functions are given by
\begin{eqnarray}\label{Kmn}
K^{\mu\nu}(Q)=2\sum_K\big(\Gamma^{\mu}(K+Q,K)G(K+Q)\nn\\
\times\gamma^{\nu}(K,K+Q)G(K)\big)+\frac{n}{m}h^{\mu\nu},
\end{eqnarray}
where $h^{\mu\nu}=-\eta^{\mu\nu}(1-\eta^{\nu0})$ is the diamagnetic current contribution, $\gamma^\mu(K+Q,K)=(1,\frac{\mathbf{k}+\frac{\mathbf{q}}{2}}{m})$ is the bare EM interaction vertex, and $\Gamma^\mu(K+Q,K)$ is the full interaction vertex to be determined later. The density-density response function is the ``00'' component of the tensor $K^{\mu\nu}$.
The density susceptibility and dynamical structure factors are
\begin{align}
&\chi''(\omega,\vq)=-\frac{1}{\pi}\textrm{Im}K^{00}(\omega,\vq),\nn\\
&S(\omega,\mathbf{q})=-\frac{1}{\pi}\cot\frac{\omega}{2T}\textrm{Im}K^{00}(\omega,\mathbf{q}).
\end{align}
Note when $\omega$ is large or $T$ is low, there is almost no difference between $\chi''$ and $S$ since the factor $\cot\frac{\omega}{2T}$ is almost 1.

In a gauge invariant linear response theory, the perturbed current must also be conserved, i.e. $q_\mu \delta  J^\mu(Q)=0$, in the whole BCS-BEC crossover regime at general temperature. This can be guaranteed by the Ward-Takahashi identity (WI) of the EM response function $q_\mu K^{\mu\nu}(Q)=0$.
These WIs satisfied by the response functions can be further inferred from the WIs satisfied by the interaction vertices
%This can be achieved by constructing a fully gauge invariant interaction vertex which satisfies the WI as the bare vertex as follows
\begin{align}
q_{\mu}\gamma^{\mu}(K+Q,K)&=G^{-1}_0(K+Q)-G^{-1}_0(K),\label{WIb}\\
q_{\mu}\Gamma^{\mu}(K+Q,K)&=G^{-1}(K+Q)-G^{-1}(K)\label{WI0},
\end{align}
and Eq.(\ref{Kmn}). The first WI can be easily verified, and the second WI indicates that the correction to the EM vertex must be in the same approximation as the self-energy effect
is included in the Green's function.
%Then is easy to verify that $q_\mu K^{\mu\nu}=0$ by using Eq.(\ref{WI0}) and the fact that $\sum_KG(K)=n$.
\begin{figure}[t]
\includegraphics[clip,width=0.45\textwidth]{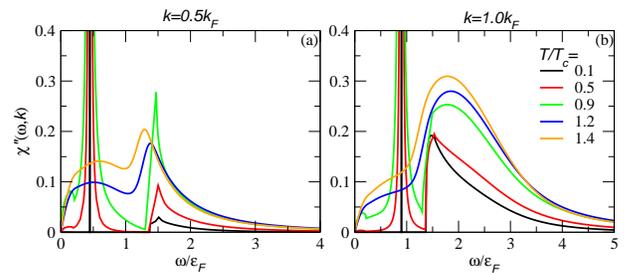}
\caption{(Color online). The dynamical density susceptibility $\chi''(\omega,k)$ (in units of $n/(2E_F)$) of a homogeneous unitary Fermi gas as a function of $\omega$, when $k=0.5k_F $ (shown in (a)) and $1.0k_F$ (shown in (b)). Here $n=k^3_F/(3\pi^2)$, $E_F=k_BT_F=\hbar^2k^2_F/(2m)$.}\label{STpgLq}
\end{figure}

A gauge invariant linear response theory for BCS mean-field theory can be developed by the integral equation formalism \cite{Nambu60}, matrix linear response theory incorporating with consistent fluctuations of order parameter \cite{KulikJLTP81,Arseev,OurJLTP13} or functional path integral approach \cite{HLYAP16}. When stronger-than-BCS attractive interaction is considered for strongly interacting Fermi gas, the gauge invariance has to be maintained in a non-trivial way due to the pair fluctuation effects. There have been some formal discussions on this subject \cite{RufusPRB16,RufusPRB16_2,HLYAP16,OurACMP15,HYPLA17}.
 %In the rest of this section, we will explain the construction of the gauge invariant full vertex $\Gamma^\mu$ which satisfies Eq.(\ref{WI0}). Such a gauge invariant vertex for BCS mean-field theory is discussed in details in Ref.\cite{OurJLTP13}. In BCS-BEC crossover, the pair fluctuation effects make full vertex more non-trivial.
A general principle to find a gauge invariant interaction vertex is the same as proving WI in quantum field theory: inserting the bare EM vertex to the self-energy diagram in all possible ways. For the pseudogap self-energy associated with the non-condensed fermion pairs, these insertions give rise to the Maki-Thompson(MT) diagram related to pseudogap, denoted by MT$_\textrm{pg}$, and the two different Aslamazov-Larkin (AL$_{1,2}$) diagrams. For the BCS self-energy associated with condensed fermion pairs, these insertions give rise to Maki-Thompson diagram related to superfluid, denoted by MT$_\textrm{sc}$, and also a contribution related to collective modes, denoted by Coll$_\textrm{sc}$ of which we haven't found a diagram representation \cite{OurACMP15}. In summary, the full vertex is now given by
\begin{align}
&\quad\Gamma^{\mu}(P+Q,P)=\gamma^{\mu}(P+Q,P)\nn\\
&+\textrm{MT}^{\mu}_{\textrm{pg}}(P+Q,P)+\textrm{AL}^{\mu}_{1}(P+Q,P)+\textrm{AL}^{\mu}_{2}(P+Q,P)\nn\\
&+\textrm{MT}^{\mu}_{\textrm{sc}}(P+Q,P)+\textrm{Coll}^{\mu}_{\textrm{sc}}(P+Q,P).\label{V1}
\end{align}
Here the MT$_\textrm{sc}$ diagram is obtained by inserting the EM vertex to the bare propagator of the superfluid self-energy
\begin{align}
&\quad\textrm{MT}_\textrm{sc}^\mu(P+Q,P)\nn\\
&=-\Delta^2_{\textrm{sc}}G_0(-P)\gamma^{\mu}(-P,-P-Q)G_0(-P-Q)\label{MTSC}
\end{align}
Similarly, the MT$_\textrm{pg}$ diagram is obtained by inserting the EM vertex to the bare propagator of the pseudogap self-energy
\begin{eqnarray}\label{MTmu}
\textrm{MT}^{\mu}_{\textrm{pg}}(P+Q,P)=\sum_Kt_{\textrm{pg}}(K)G_{0}(K-P)\nn\\
\times\gamma^{\mu}(K-P,K-P-Q)G_{0}(K-P-Q).
\end{eqnarray}
The two AL diagrams can be obtained by inserting EM vertex to the pair propagator
\begin{eqnarray}
&&\textrm{AL}^{\mu}_{1}(P+Q,P)=-\sum_{K,L}t_{\textrm{pg}}(K)t_{\textrm{pg}}(K+Q)G_{0}(K-P)\nn\\
&&\times G(K-L)G_{0}(L+Q)\gamma^{\mu}(L+Q,L)G_{0}(L),\nonumber\\
&&\textrm{AL}^{\mu}_{2}(P+Q,P)=-\sum_{K,L}t_{\textrm{pg}}(K)t_{\textrm{pg}}(K+Q)G_{0}(K-P)\nn\\
&&\times G_{0}(K-L)G(L+Q)\Gamma^{\mu}(L+Q,L)G(L).\label{ALmu}
\end{eqnarray}
Note that due to the asymmetric choice of $t$-matrix, we have two different AL terms in $G_0G$ theory, while there is only on AL term in the Nozieres Schmitt-Rink(NSR) crossover theory \cite{NSR}.
The last one is the collective-mode diagram
\begin{align}
\textrm{Coll}_\textrm{sc}^\mu(K+Q,K)&=\Delta_{\textrm{sc}}\Pi^{\mu}(Q)G_0(-K-Q)\nn\\
&+\Delta_{\textrm{sc}}\bar{\Pi}^{\mu}(Q)G_0(-K),\label{CSC}
\end{align}
where $\Pi^{\mu}=\Pi_1^{\mu}+i\Pi_2^{\mu}$ and $\bar{\Pi}^{\mu}=\Pi_1^{\mu}-i\Pi_2^{\mu}$, with
\begin{eqnarray}\label{tmp4}
\Pi^{\mu}_1=\frac{\left|\begin{array}{cc}
Q_{21} & Q^{\mu}_{31}\\
\tilde{Q}_{22} & Q^{\mu}_{32}
\end{array}\right|}
{\left|\begin{array}{cc}
\tilde{Q}_{11} & Q_{12}\\
Q_{21} & \tilde{Q}_{22}
\end{array}\right|},\quad
\Pi^{\mu}_2=\frac{\left|\begin{array}{cc}
\tilde{Q}_{11} & Q^{\mu}_{31}\\
Q_{21} & Q^{\mu}_{32}
\end{array}\right|}
{\left|\begin{array}{cc}
\tilde{Q}_{11} & Q_{12}\\
Q_{21} & \tilde{Q}_{22}
\end{array}\right|},
\end{eqnarray}
Here $\tilde{Q}_{11}=Q_{11}+\frac{2}{g}$, $\tilde{Q}_{22}=Q_{22}+\frac{2}{g}$ and the response functions $Q_{ij}$ are given in Appendix.\ref{appb}. This diagram only appears in the superfluid phase (below $T_c$) where the symmetry is spontaneously broken. In the strict mean field theory, this diagram can be obtained by inserting the EM vertex into the gap equation at all possible places \cite{HYPLA17}.

\begin{figure}[t]
\includegraphics[clip,width=0.45\textwidth]{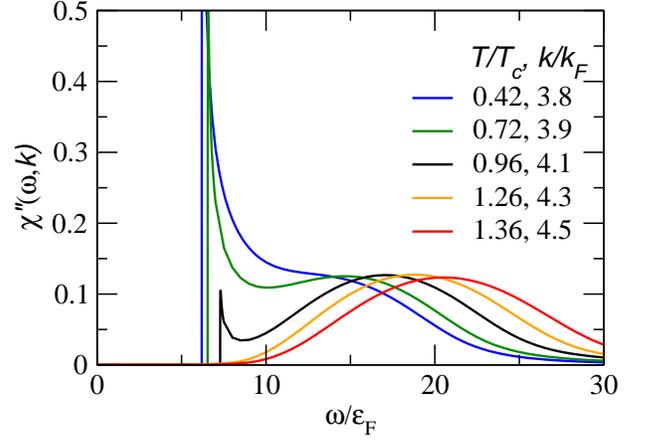}
\caption{(Color online). The dynamical density susceptibility $\chi''(\omega,k)$ (in units of $n/(2E_F)$) of a homogeneous unitary Fermi gas as a function of $\omega$ when the momentum transfer is relatively large. These 5 different colored curves correspond to 5 different sets of values for $T/T_c$ and $k/k_F$, which are labeled in the figure.}\label{STpg}
\end{figure}

The full EM vertex given by Eq.(\ref{V1}) satisfies the Ward identity (\ref{WI0}) $q_{\mu}\Gamma^{\mu}(K+Q,K)=G^{-1}(K+Q)-G^{-1}(K)$. A proof by combining Eqs.(\ref{TWI}), (\ref{MWI}) and (\ref{CWI}) can be found in Appendix.\ref{appc}
The full density-density response function is now expressed by
\begin{align}
&\quad K^{00}=Q^{00}_{33}+\delta K^{00}_{33}\nn\\
&\delta K^{00}_{33}=-\frac{\tilde{Q}_{11}Q^{0}_{32}Q^{0}_{23}+\tilde{Q}_{22}Q^{0}_{31}Q^{0}_{13}}
{\tilde{Q}_{11}\tilde{Q}_{22}-Q_{12}Q_{21}}\nn\\
&\qquad+\frac{Q_{12}Q^{0}_{31}Q^{0}_{23}+Q_{21}Q^{0}_{32}Q^{0}_{13}}
{\tilde{Q}_{11}\tilde{Q}_{22}-Q_{12}Q_{21}},
\end{align}
where the expressions of the response functions are given in Appendix.\ref{appb}. $\delta K^{00}_{33}$ comes from the collective-mode contribution. Hence it vanishes above $T_c$ since there is no symmetry breaking there, and the poles of it determines the dispersion of collective modes (Nambu-Goldstone modes). The WI ensures that the structure factor satisfies the $f$-sum rule
\begin{align}
\int_{-\infty}^{+\infty}d\omega\omega \chi''(\omega,\mathbf{q})=\frac{nq^2}{2m}.
\end{align}

\section{numerical results and discussions}
\label{sec-num}

%\subsection{results for large incoming momentum}

\begin{figure}[t]
\includegraphics[clip,width=0.4\textwidth]{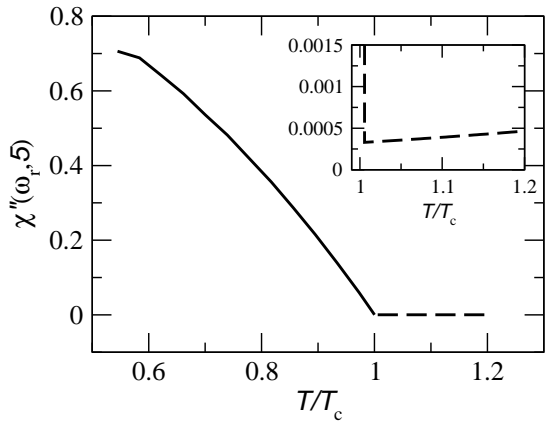}
\caption{The dynamical density susceptibility $\chi''$ (in units of $n/(2E_F)$) as a function of $T/T_c$ for $k/k_F=5$ and $\omega=\omega_0$, where $\omega_0$ is peak location below $T_c$ and $\omega_0=\omega_r/2$ above $T_c$. Panel (a) and (b) are calculated from the mean field theory and pair fluctuation theory respectively. The insets denote the blows up of the $\chi''$ curves around $T_c$.}\label{STm}
\end{figure}

Now we present the numerical results of our theory and compare them with the experiments. We first calculate the dynamic density structure factor of uniform unitary Fermi gases from low to high temperatures with relatively small transfer momentum. The results are shown in Figure \ref{STpgLq} where $k=0.5k_F$ in (a) and $1.0k_F$ in (b). Here $k_F$ is the Fermi momentum of a noninteracting Fermi gas with
the same density. In our $G_0G$ theory, the critical temperature is $T_c\simeq 0.26T_F$ at unitary limit, which is about 40\% higher than the experiment value $T_c/T_F=0.167$. This difference is due to the inaccurate estimation of the mean field background, which is also known from other thermodynamic quantities. However, we can expect that the dynamical density susceptibility is generally insensitive to this background. When $T<T_c$, there is clearly a peak associated with the Nambu-Goldstone mode due to the spontaneous breaking of the U(1) symmetry in the superfluid phase, which gives rise to a collective motion of the condensate. When the transferring momentum is small enough, this corresponds to the well known Bogoliubov-Anderson mode.
In the numerics, the imaginary part of $\omega$, which comes from the complex continuation $i\omega_n\rightarrow\omega+i0^+$, in the response functions is assumed to be infinitesimally small below $T_c$, which can be understood by assuming that the lifetime of the pairs is long enough below $T_c$. This makes the peaks in Figure \ref{STpgLq} much sharper than those of the experiments. There are also two single-particle continua, one corresponds to the quasiparticle scattering process with $\omega=\pm(E^+_\mathbf{k}-E^-_\mathbf{k})$ where $E^\pm_\mathbf{k}=E_{\mathbf{k}\pm\frac{\mathbf{q}}{2}}$ are quasiparticle dispersions, another corresponds to the pair breaking process with $\omega=E^+_\mathbf{k}+E^-_\mathbf{k}>2\Delta$. For details please refer to the expressions of the response function list in Appendix.\ref{appb}. When the transfer frequency is low enough the former continuum branch becomes quite apparent in the curves denoted by the $T=0.1T_c$, $0.5T_c$. This agrees with the recent experiment\cite{ValeNP17}. When $T>T_c$, the collective-mode peak disappear due to the restoration of the symmetry, and the continua join together. As we increase the transferring momentum $k$, the collective-mode peak approaches the pair-breaking continuum (see Figure \ref{STpgLq}(b)). Finally they will merge together is $k$ is large enough.

In Figure \ref{STpg}, the dynamical density susceptibility $\chi''$ are plotted as a function of $\omega$ for large transferring momenta. As we discussed before, all continua and peak merge together. The five different curves correspond to $T/T_c$=0.42, 0.72, 0.96, 1.26, 1.36 and $k/k_F$=3.8, 3.9, 4.1, 4.3, 4.5 respectively. Here we choose the relative temperature $T/T_c$ in figure \ref{STpg} according to the experimental results in Ref.\cite{Vale14}. Below $T_c$, $\chi''$ displays a sharp peak roughly at $1/2$ of the so-called recoil frequency $\omega_r=k^2/(2m)$ when $k$ is large enough such that $\omega_r/2>2\Delta$. This sharp peak of $\chi''$ clearly corresponds to the scattering of the pairs from momentum conservation consideration. Since the lifetime of the pair is assumed to be long enough below $T_c$ as we discussed previously, the peaks in Figure \ref{STpg} are also much sharper than those of the experiments.
 Above $T_c$, this sharp peak disappears again and only a very broad continuum roughly centered around $\omega_r$, which corresponds to the pair breaking ($\omega>2\Delta$) and quasiparticle scattering.
Here we have assumed that the recoil frequency in the response functions has a nonzero imaginary part $\omega+i\frac{1}{\tau}$ above $T_c$ after implementing analytical continuation, since the non-condensed pairs have a finite lifetime. The lifetime is chosen approximately as $\tau\simeq 10E^{-1}_F$ to ensure a suitable width of the peak.
This qualitatively agrees with what has been observed in the experiments \cite{Vale14}.

In Figure \ref{STm}, we plot the dynamical density susceptibility $\chi''$ as a function of $T/T_c$ for fixed large momentum $k/k_F=5$ and fixed frequency $\omega=\omega_0$, where $\omega_0$ is the location of the peaks below $T_c$ and $\omega_0=\omega_r/2$ above $T_c$. Below $T_c$, the peak location is not exactly at $\omega_r/2$. Only when the momentum transfer is large, the peak location is close to $\omega_r/2$ since the characteristic length is much less than the size of condensed pairs. Hence we choose to plot the value of $\chi''$ at the top of the peak, which makes the curve blow $T_c$ not very smooth. Above $T_c$, there is no sharp peak any more, thus we simply plot the value of $\chi''$ at $\omega=\omega_r/2$. One can clearly see that the temperature dependence of $\chi''(\omega_0)$ is roughly linear both below and above $T_c$, but with quite different slopes. Above $T_c$, $\chi''(\omega_0)$ is almost flat, while below $T_c$, $\chi''(\omega_0)$
increases much faster with decreasing $T$. This indicates that the appearing of condensed pairs greatly increases the pair scattering. Therefore, $\chi''(\omega_0)$ can serve as an order parameter of the condensate in certain sense. In the insets, we show the blow up of the details around $T_c$. One can see the flat curve above $T_c$ actually increases slightly with the increasing temperature. This trend is the same as the ideal gas.

%\subsection{results for small incoming momentum}

\section{Conclusion}
In summary, we construct a manifestly gauge invariant linear response theory for a strongly correlated Fermi gas, from which the dynamic density structure factor can be studied near the Feshbach resonance. Our numerical results qualitatively agrees with the known experimental results.

%{\bf INCLUDE SMALL q CASES?}

\textit{Acknowledgment ---}
Yan He is supported by NSFC under grant No.11404228. Hao Guo is supported by NSFC under grant No.11674051.

\appendix

\section{Response Functions with Pair Fluctuation}\label{appb}
The response functions are given by
\begin{widetext}
\begin{align}
Q_{11}(\omega,\vq)&=\sumk\Big[(1+\frac{\xik^+\xik^--\Delta^2}{\Ek^+\Ek^-})
\frac{\Ek^++\Ek^-}{\omega^2-(\Ek^++\Ek^-)^2}[1-f(\Ek^+)-f(\Ek^-)]\nn\\
&\qquad-(1-\frac{\xik^+\xik^--\Delta^2}{\Ek^+\Ek^-})
\frac{\Ek^+-\Ek^-}{\omega^2-(\Ek^+-\Ek^-)^2}[f(\Ek^+)-f(\Ek^-)]\Big]\\
Q_{12}(\omega,\vq)&=Q_{21}(\omega,\vq)\nn\\
&=-i\omega\sumk\Big[(\frac{\xik^+}{\Ek^+}+\frac{\xik^-}{\Ek^-})\frac{1-f(\Ek^+)-f(\Ek^-)}{\omega^2-(\Ek^++\Ek^-)^2}
+(\frac{\xik^+}{\Ek^+}-\frac{\xik^-}{\Ek^-})\frac{f(\Ek^+)-f(\Ek^-)}{\omega^2-(\Ek^++\Ek^-)^2}\Big]\\
Q_{13}^0(\omega,\vq)&=Q_{31}^0(\omega,\vq)\nn\\
&=\Delta_\textrm{sc}\sumk\frac{\xik^++\xik^-}{\Ek^+\Ek^-}\Big[\frac{(\Ek^++\Ek^-)[1-f(\Ek^+)-f(\Ek^-)]}{\omega^2-(\Ek^++\Ek^-)^2}
+\frac{(\Ek^+-\Ek^-)[f(\Ek^+)-f(\Ek^-)]}{\omega^2-(\Ek^+-\Ek^-)^2}\Big]\\
Q_{13}^j(\omega,\vq)&=Q_{31}^j(\omega,\vq)\nn\\
&=\Delta_\textrm{sc}\sumk\frac{k^j}{m}\frac{\omega}{\Ek^+\Ek^-}\Big[\frac{(\Ek^++\Ek^-)[1-f(\Ek^+)-f(\Ek^-)]}{\omega^2-(\Ek^++\Ek^-)^2}
+\frac{(\Ek^+-\Ek^-)[f(\Ek^+)-f(\Ek^-)]}{\omega^2-(\Ek^+-\Ek^-)^2}\Big]\\
Q_{22}(\omega,\vq)&=\sumk\Big[(1+\frac{\xik^+\xik^-+\Delta^2}{\Ek^+\Ek^-})
\frac{\Ek^++\Ek^-}{\omega^2-(\Ek^++\Ek^-)^2}[1-f(\Ek^+)-f(\Ek^-)]\nn\\
&\qquad-(1-\frac{\xik^+\xik^-+\Delta^2}{\Ek^+\Ek^-})
\frac{\Ek^+-\Ek^-}{\omega^2-(\Ek^+-\Ek^-)^2}[f(\Ek^+)-f(\Ek^-)]\Big]\\
Q_{23}^0(\omega,\vq)&=-Q_{32}^0(\omega,\vq)\nn\\
&=i\Delta_\textrm{sc}\sumk\frac{\omega}{\Ek^+\Ek^-}\Big[\frac{(\Ek^++\Ek^-)[1-f(\Ek^+)-f(\Ek^-)]}{\omega^2-(\Ek^++\Ek^-)^2}
+\frac{(\Ek^+-\Ek^-)[f(\Ek^+)-f(\Ek^-)]}{\omega^2-(\Ek^+-\Ek^-)^2}\Big]\\
Q_{23}^j(\omega,\vq)&=-Q_{32}^j(\omega,\vq)\nn\\
&=i\Delta_\textrm{sc}\sumk\frac{k^j}{m}\frac{\xik^+-\xi^-}{\Ek^+\Ek^-}\Big[\frac{(\Ek^++\Ek^-)[1-f(\Ek^+)-f(\Ek^-)]}{\omega^2-(\Ek^++\Ek^-)^2}
+\frac{(\Ek^+-\Ek^-)[f(\Ek^+)-f(\Ek^-)]}{\omega^2-(\Ek^+-\Ek^-)^2}\Big]\\
Q_{33}^{00}(\omega,\vq)&=\sumk\Big[(1-\frac{\xik^+\xik^--\Delta^2_\textrm{sc}+\Delta^2_\textrm{pg}}{\Ek^+\Ek^-})
\frac{\Ek^++\Ek^-}{\omega^2-(\Ek^++\Ek^-)^2}[1-f(\Ek^+)-f(\Ek^-)]\nn\\
&\qquad-(1+\frac{\xik^+\xik^--\Delta^2_\textrm{sc}+\Delta^2_\textrm{pg}}{\Ek^+\Ek^-})
\frac{\Ek^+-\Ek^-}{\omega^2-(\Ek^+-\Ek^-)^2}[f(\Ek^+)-f(\Ek^-)]\Big]\\
Q_{33}^{ij}(\omega,\vq)&=\sumk\frac{k^ik^j}{m^2}\Big[(1+\frac{\xik^+\xik^-+\Delta^2_\textrm{sc}-\Delta^2_\textrm{pg}}{\Ek^+\Ek^-})
\frac{\Ek^++\Ek^-}{\omega^2-(\Ek^++\Ek^-)^2}[1-f(\Ek^+)-f(\Ek^-)]\nn\\
&\qquad-(1-\frac{\xik^+\xik^-+\Delta^2_\textrm{sc}-\Delta^2_\textrm{pg}}{\Ek^+\Ek^-})
\frac{\Ek^+-\Ek^-}{\omega^2-(\Ek^+-\Ek^-)^2}[f(\Ek^+)-f(\Ek^-)]\Big]\\
Q_{33}^{0j}(\omega,\vq)&=Q_{21}^{j0}(\omega,\vq)\nn\\
&=\omega\sumk\frac{k^j}{m}\Big[(\frac{\xik^+}{\Ek^+}-\frac{\xik^-}{\Ek^-})\frac{1-f(\Ek^+)-f(\Ek^-)}{\omega^2-(\Ek^++\Ek^-)^2}
+(\frac{\xik^+}{\Ek^+}+\frac{\xik^-}{\Ek^-})\frac{f(\Ek^+)-f(\Ek^-)}{\omega^2-(\Ek^++\Ek^-)^2}\Big]
\end{align}
Here $\xi^\pm_\mathbf{k}=\xi_{\mathbf{k}\pm\frac{\mathbf{q}}{2}}$ and $E^\pm_\mathbf{k}=E_{\mathbf{k}\pm\frac{\mathbf{q}}{2}}$.
\end{widetext}

\section{Gauge Invariance of the Linear Response Theory}\label{appc}
By applying WI (\ref{WIb}), we find
\begin{eqnarray}
& &q_{\mu}\textrm{MT}^{\mu}_{\textrm{sc}}(P+Q,P)=\Sigma_{\textrm{sc}}(P+Q)-\Sigma_{\textrm{sc}}(P),\label{MWI}\\
& &q_{\mu}\textrm{MT}^{\mu}_{\textrm{pg}}(P+Q,P)=\Sigma_{\textrm{pg}}(P+Q)-\Sigma_{\textrm{pg}}(P).\label{MTpgWI}
\end{eqnarray}
If we try to combine these two relations with the ``bare'' WI (\ref{WIb}) to prove the ``full'' WI (\ref{WI0}), it seems we have ``wrong'' minus signs for self-energies. In fact, the two AL diagrams and collective-mode diagram serve to ``flip'' this extra minus sign.
It can be easily shown that the MT and AL diagrams satisfy a relation
\begin{eqnarray}
q_{\mu}\big[\frac{1}{2}\textrm{AL}^{\mu}_{1}(P+Q,P)+\frac{1}{2}\textrm{AL}^{\mu}_{2}(P+Q,P)\nn\\
+\textrm{MT}^{\mu}_{\textrm{pg}}(P+Q,P)\big]=0.
\end{eqnarray}
Together with identity (\ref{MTpgWI}), this relation implies
\begin{eqnarray}\label{TWI}
&&q_{\mu}\big[\textrm{AL}^{\mu}_{1}(P+Q,P)+\textrm{AL}^{\mu}_{2}(P+Q,P)]\nn\\
&=&2\Sigma_{\textrm{pg}}(P)-2\Sigma_{\textrm{pg}}(P+Q).
\end{eqnarray}
It can be proved that
\begin{align}
q_\mu\Pi^{\mu}(Q)&=2\Delta_\textrm{sc},\notag\\
q_\mu\bar{\Pi}^{\mu}(Q)&=-2\Delta_\textrm{sc},
 \end{align}
which in turn gives
\begin{eqnarray}\label{CWI}
q_{\mu}\textrm{Coll}^{\mu}_{\textrm{sc}}(P+Q,P)=2\Sigma_{\textrm{sc}}(P)-2\Sigma_{\textrm{sc}}(P+Q).
\end{eqnarray}
Applying Eqs.(\ref{WIb}), (\ref{MWI}), (\ref{MTpgWI}), (\ref{TWI}) and  (\ref{CWI}) together, we can finally get the WI (\ref{WI0})
\begin{align}
q_{\mu}\Gamma^{\mu}(K+Q,K)&=G^{-1}(K+Q)-G^{-1}(K).
\end{align}
Therefore the linear response theory is gauge invariant. The WI guarantees the validity of the $f$-sum rule.

%\bibliographystyle{apsrev}
%\bibliography{Review,Review1,Review2}

\end{document}